\documentclass{article}
\usepackage{spconf,amsmath,graphicx}
\usepackage{acronym}

\usepackage[ruled,vlined]{algorithm2e}
\SetKwInput{KwInput}{Input}
\usepackage{enumitem}
\usepackage{tikz}
\usepackage{adjustbox}
\usepackage{url,times,amssymb}
\usepackage{amsopn}
\usepackage{balance}
\usepackage{cite}
\usepackage{multirow}
\usepackage{hhline}
\usepackage{booktabs}
\usepackage{array}
\usepackage{url}
\usepackage{setspace}
\usepackage{epstopdf}
\usepackage{mathtools}

\acrodef{DNN}{deep neural network}
\acrodef{ATF}{acoustic transfer function}
\acrodef{STFT}{short time Fourier transform}
\acrodef{RT}{Reverberation Time}
\acrodef{SNR}{Signal To Noise Ratio}
\acrodef{FC}{fully-connected}
\acrodef{ADAM}{Adaptive Moment Estimation}
\acrodef{ReLU}{Rectified Linear Unit}
\acrodef{NN}{neural network}
\acrodef{CNN}{convolutional neural network}
\acrodef{TF}{time frequency}
\acrodef{PESQ}{Perceptual Evaluation of Speech Quality}
\acrodef{RIR}{room impulse response}
\acrodef{S2S}{spectrum to spectrum}
\acrodef{CD}{ceprtrum distance}
\acrodef{LLR}{log likelihood ratio}
\acrodef{RI}{real and imaginary}
\acrodef{SI-SDR}{scale-invariant signal to distortion ratio}
\acrodef{SA}{self-attention}
\acrodef{MSE}{mean square error}
\acrodef{STD}{standard deviation}
\acrodef{SRMR}{speech-to reverberation modulation energy ratio}
\acrodef{CP}{clean phase}
\acrodef{NP}{noisy phase}
\acrodef{EP}{enhanced phase}
\acrodef{RI2RI}{real and imaginary to real and imaginary}
\acrodef{GAN}{generative adversarial network}
\acrodef{RNN}{recurrent neural network}
\acrodef{TF}{time-frequency}
\acrodef{WPE}{weighted prediction error}
\acrodef{SI-SDR}{scale-invariant signal to distortion ratio}
\acrodef{BSS}{blind source separation}
\acrodef{SOTA}{state-of-the-art}

\title{Magnitude or Phase? A Two Stage Algorithm for Dereverberation}

\name{Ayal Schwartz$^1$, Sharon Gannot$^2$ and Shlomo E.~Chazan$^1$}
\address{$^1$ OriginAI, Ramat-Gan, Israel\\
$^2$  Faculty of Engineering, Bar-Ilan University, Ramat-Gan, Israel
}

\begin{document}
%
\maketitle
\begin{abstract}


In this work we present a new single-microphone speech dereverberation algorithm. First, a performance analysis is presented to interpret that algorithms focused on improving solely magnitude or phase are not good enough. Furthermore, we demonstrate that few objective measurements have high correlation with the clean magnitude while others with the clean phase.  Consequently ,we propose a new architecture which consists of two sub-models, each of which is responsible for a different task. 
The first model  estimates the clean magnitude given the noisy input. The enhanced magnitude together with the noisy-input phase are then used as inputs to  the second model to estimate the real and imaginary portions of the dereverberated signal. A training scheme including pre-training and fine-tuning is presented in the paper.
We evaluate our proposed approach using data from the REVERB challenge and compare our results to other methods. We demonstrate consistent improvements in all measures, which can be attributed to the improved estimates of both the magnitude and the phase.

\end{abstract}
\begin{keywords}
dereverberation, decoupling model 
\end{keywords}

\label{sec:intro}

\section{Introduction}
Single-channel speech dereverberation has been a field of extensive research for many years, and is still regarded as a very challenging task~\cite{kinoshita2016summary}. One of the leading algorithms is the long-term linear prediction method with \ac{WPE} proposed in~\cite{delcroix2014linear}.

In recent years, \ac{DNN}-based  models were utilized to deal with this challenge. The majority of these methods attempts to enhance the magnitude of the noisy and reveberant \ac{STFT}~\cite{ernst2018speech, xiao2016speech, ribas2019deep}. In these approaches, the enhanced magnitude is combined with the noisy phase and then inverse-transformed to the time-domain. The noisy phase is incompatible with the enhanced magnitude, resulting in noticeable artifacts in the enhanced signal. 
A U-Net \ac{GAN} architecture was proposed in~\cite{ernst2018speech} for estimating the \ac{STFT} magnitude. The GAN block contributed additional improvement. However, as the noisy phase is  used, speech distortion is still audible. 

Recently, algorithms which take the phase into account were developed. 
The Griffin-Lim method~\cite{griffin1984signal} was used in~\cite{zhao2020monaural} to enhance the noisy phase. Furthermore, the HiFi-GAN~\cite{kong2020hifi} was used to reconstruct the raw samples of the clean signal given the noisy magnitude in~\cite{su2020hifi}.
Unfortunately this approach tends to perform poorly when introduced with unfamiliar data. Decomposing the noisy \ac{STFT} to Real and Imaginary (RI) components was proposed in~\cite{wang2020deep,li2021importance}. In this approach, the goal is to estimate the clean RI of the clean signal rather than only its magnitude. Usually, this model is trained with \ac{MSE} objective, which might not be the best loss for this task. 

\begin{figure}[t!]
\centering
\includegraphics[width=0.5\textwidth]{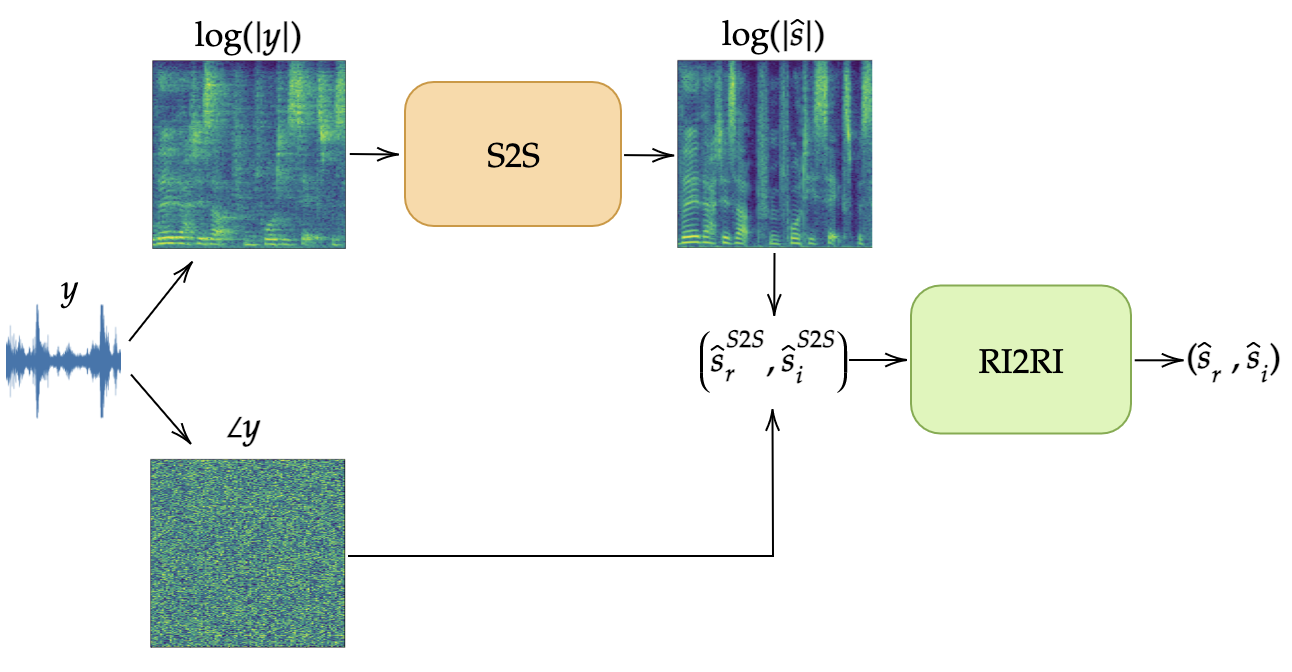}
\caption{The proposed magnitude and phase decoupling model.}
\label{fig:proposed}
\end{figure}
\sloppy
Recently, the \ac{SI-SDR} loss~\cite{le2019sdr} was introduced and found useful in many  downstream audio tasks such as \ac{BSS}~\cite{luo2019conv,chazan2021single} and  speech enhancement~\cite{hu2020dccrn,kolbaek2020loss}. Since the \ac{SI-SDR} is applied in the time-domain, it has a tendency to enhance the noisy phase together with the noisy magnitude, which is also estimated by traditional methods. This observation was reported in~\cite{wang2020deep}. 
A two-stage algorithm for speech denoising was presented in~\cite{li2021icassp}. 

In the current contribution, we present a decoupled dereverberation approach. First, a fully convolution model (U-net architecture) with self-attention (SA) units is utilized to estimate the clean magnitude given the noisy signal. Then, the enhanced magnitude together with the noisy phase are transformed into \ac{RI}, which are then used to estimate the  \ac{RI} parts of the clean signal. This model is trained with  the \ac{SI-SDR} loss which was found to improve the phase.
Experiments show that the proposed approach outperforms both magnitude-based and RI2RI-based models.

\section{Problem Formulation}
\label{sec:problem}

Let $y(t)$ be the received microphone signal in a noisy and reveberant environment:
\begin{equation}
    y(t) = x(t) + n(t),\quad t=1,\ldots,T
    \label{eq:noisy_time}
\end{equation}

\begin{equation}
    x(t) = \{s\ast h\}(t)
    \label{eq:conv}
\end{equation}
where $s(t)$, $h(t)$, $n(t)$ are the discrete-time desired speech signal, the \ac{RIR} relating the speaker and the microphone, and the additive background noise, respectively. $t$ is the discrete time index, and $T$ is the number of available samples. In the \ac{STFT} domain, \eqref{eq:noisy_time} is given by,
\begin{equation}
    y(l,k) = x(l,k) + n(l,k),
    \label{eq:noisy_tf}
\end{equation}
where $y(l,k)$, $x(l,k)$ and $n(l,k)$ denote the \ac{STFT} representations of $y(t),x(t)$, and $n(t)$, respectively, and l and k denote the time and frequency indices, respectively. 
The goal of this work is to estimate the clean signal $s(t)$ given the noisy and reverberated signal $y(t)$.

        



\begin{table}

 \caption{ Performance of different phases / magnitude combinations on SimData of REVERB challenge evaluation set. It is evident that the first three measurements are more affected by the magnitude while the \ac{SI-SDR} is more affected by the phase.}
\centering
\resizebox{\linewidth}{!}{%
 \begin{tabular}{l l r r r r}
 \midrule
 Magnitude & Phase & \textbf{LLR} $\downarrow$ & \textbf{CD}  $\downarrow$ & \textbf{PESQ}  $\uparrow$ & \textbf{SI-SDR}  $\uparrow$
\\ \midrule
Noisy & Noisy  & 0.58 & 3.97 & 1.48 & -10.4 \\
Noisy & Clean   & 0.52 & 3.81 & 1.59 &\textbf{ 5.11} \\
Clean & Noisy   & \textbf{0.04} & \textbf{1} & \textbf{3.35} & -8.9 \\
\midrule 
\end{tabular}}
\label{table:1}
\end{table}

\section{Magnitude vs phase analysis}
\label{sec:mag_vs_phase}


To evaluate the consequences of improving only the magnitude or the phase components of the speech, we conducted an analysis experiment. 
In this experiment, we used the simulated signals from the REVERB challenge~\cite{kinoshita2016summary}. Three combination variants were tested: 1)  noisy magnitude with  noisy phase, 2)  noisy magnitude with  clean phase, and 3)  clean magnitude with  noisy phase. To evaluate these variants we calculated the following objective measures, the \ac{CD}, the \ac{LLR}, the \ac{PESQ}, and the \ac{SI-SDR}. 

Table~\ref{table:1} presents the average metrics for the different variants. We found that the variant with the clean magnitude and the noisy phase dramatically improves  the first three quality measures while barley improves the \ac{SI-SDR}. In contrast, the variant with the noisy magnitude and the clean phase improves the \ac{SI-SDR} while  only marginally improving the \ac{CD}, the \ac{LLR} and the \ac{PESQ}. 
It is therefore evident that the first three measures are more affected by the magnitude, while the latter is more affected by the phase. To the best of our knowledge, this is the first analysis that addresses this point.

As mentioned in the introduction, working solely with either the magnitude or with the phase does not guarantee improvement to the other. Hence, we aim to find a better operating point that is beneficial for both the magnitude and the phase.

\section{Proposed Algorithm} \label{sec:Algorithm}

\subsection{Network architecture}




A block diagram of the proposed model is depicted in Fig.~\ref{fig:proposed}. 
A two-stage architecture for speech dereverberation is introduced. 
Our model is comprised of two sub-models, namely the \ac{S2S} and the \ac{RI2RI} blocks. The \ac{S2S} block uses as input features the log-magnitude of the noisy signal and is trained to estimate the log-magnitude of the clean speech. 
The \ac{RI2RI} block is trained to estimate the \ac{RI} of the clean signal given the enhanced log-magnitude from the \ac{S2S} model and the phase of the noisy signal as input features. The first model emphasizes the harmonic structure of the clean signal and simultaneously, reduces most of the reverberation and background noise. The second model mainly improves the noisy phase, given the enhanced magnitude (from the first stage) and noisy phase.

The \ac{S2S} block is implemented with a U-net architecture with \ac{SA} units on the bottleneck-latent layer. The network is constructed with an encoder and a decoder with skip-connection connecting them layer-wise.  The down-sampling and up-sampling operation is only applied  to the frequency axis to preserve the temporal information which is utilized by the \ac{SA} layer.  The output activation function of the \ac{S2S} model is  `tanh' with learned amplification gain estimated from the latent space to restore the clean log-magnitude range. 

The \ac{RI2RI} model is similarly implemented, but without the \ac{SA} layer, which degraded the performance in our study. This might be explained due to
the random characteristics of the phase.

\subsection{Training objectives}\label{sec:pre_train}

To train this a complex structure, a pre-training stage is required. Since  the \ac{S2S} and the {RI2RI} blocks are designed to accomplish different tasks, each of them is trained with its own objective.

The \ac{S2S} block is trained on mapping the noisy log-magnitude to the clean log-magnitude in an image-to-image manner. We therefore train the block with the   \ac{MSE} loss function:

\begin{equation}
    \mathrm{Loss}_{\mathrm{S2S}}= \frac{1}{L K}\sum_{l=0}^{L-1}\sum_{k=0}^{K-1}{(\text{log}|\hat{s}(l,k)|-\text{log}|s(l,k)| )}^2
    \label{eq:mse_loss}
\end{equation}
where $\hat{s}(l,k)$ and ${s(l,k)}$ refer to the enhanced and clean complex \ac{STFT} images, respectively.  This \ac{S2S} block was separately pre-trained with~\eqref{eq:mse_loss}. 

The \ac{RI2RI} model is trained to map the noisy \ac{RI} to the clean \ac{RI}. In the proposed algorithm, this block focuses on the phase enhancement, which, as shown in Sec.~\ref{sec:mag_vs_phase}, is strongly related to the \ac{SI-SDR} measure, implemented in the time domain. Therefore, the loss function of the \ac{RI2RI} model is given by:

\begin{equation}
    \mathrm{Loss}_{\mathrm{RI2RI}}=10\log_{10}\left(\frac{||\alpha \textbf{s}||^2}{||\alpha \textbf{s}-\hat{\textbf{s}}||^2}\right)
    \label{eq:sisdr_loss}
\end{equation}
with
\begin{equation}
    \alpha= \frac{\hat{\textbf{s}}^T\textbf{s}}{||\textbf{s}||^2}
    \label{eq:alpha}
\end{equation}
where $\hat{\textbf{s}}$ and ${\textbf{s}}$ are vectors referring to the entire enhanced and clean raw signals in time-domain, respectively. We note that, \eqref{eq:mse_loss} and \eqref{eq:sisdr_loss} are averaged over a mini-batch with length $N_b$. In order to direct the \ac{RI2RI} model to  focus on the phase estimation task, it was pre-trained with a synthetic data constructed with clean magnitude and noisy phase. The pre-training of this model also utilizes \eqref{eq:sisdr_loss}. 

Once the two blocks are pre-trained, a joint training is carried out to fine-tune the model. In our experiments we found that only the \ac{RI2RI} model requires the fine-tuning stage. Hence the \ac{S2S} weights were frozen in that stage. This will be further discussed in the Sec.~\ref{Ablation_study}. 



\section{Experimental Study}

\subsection{The REVERB challenge data}
The REVERB challenge corpus~\cite{kinoshita2016summary} was both used to train and to evaluate our proposed method. The training-set consists of 7138 clean utterances of 83 speakers from WSJ0 dataset.
To simulate the reverberant speech utterances with a \ac{SNR} of 20 dB,  24 measured room impulse responses and pre-recorded background noise were used. The development test set  consists of
1484 speech utterances from the WSJCAM0 dataset
that are convolved with measured RIRs  to generate reverberant speech utterances. The recorded background noise
is a stationary diffuse noise mainly caused by the air-conditioning
systems in the rooms. The  \acp{RIR} in the training and development test sets are characterized by reverberation time ($\text{RT}_{60}$) in the range from 0.2~s to 0.8~s and the distance between the source and the microphone in the range 0.5~m-2.5~m.
The room dimensions and the acoustic conditions were different for the evaluation set and the training set. 

\subsection{Pre-Processing}
The \ac{STFT} was computed using frame length of 512 samples (corresponding to 32 msec in sampling rate, $F_s$=16KHz), multiplied by a Hamming window and with an overlap of 256 samples. Due to the symmetry characteristic of the STFT only the first 257 frequency bins were used. 
The input signal was normalize by its \ac{STD} as input normalization.
SpecAugment
was used in the training phase of the \ac{S2S} model. In the RI2RI model this augmentation did not improve the results and was therefore not used. 
Finally, the input features is shape to the  model was set to $256 \times 256$ during training, while in inference it is not restricted to a specific length while the width is always the same due to constant \ac{STFT} window. 

\subsection{Compared methods and evaluation}
To evaluate our method we compared it with the model-based WPE-based algorithm~\cite{delcroix2014linear}, and four DNN-based algorithms~\cite{xiao2016speech,ernst2018speech,ribas2019deep,zhao2020monaural}. 
Furthermore, we trained three additional models. The first, is a S2S model with SA units. This model was tested once with the the noisy phase (dubbed S2S+NP) and once with the clean phase (dubbed S2S+CP). The second is an \ac{RI2RI} model trained with MSE loss (dubbed RI2RI (MSE)). The last one is an \ac{RI2RI} model trained with the \ac{SI-SDR} loss (dubbed RI2RI (SI-SDR)). The first one is a magnitude-based model, and the other two are phase-aware models, as mentioned in the introduction, that  take the entire STFT information into account in one stage.

To evaluate the performance and compare between the competing algorithms, we used the following objective measures. The \ac{PESQ}~\cite{rix2001perceptual}, the \ac{CD} and the \ac{LLR} which are known to be more correlated with the magnitude domain. In addition, we applied the frequency-weighted segmental signal-to-noise ratio ($\mathrm{SNR}_{\textrm{fw}}$), and the \ac{SRMR}. Finally, we evaluated the \ac{SI-SDR}, which is more correlated to the phase domain, as shown in Sec.~\ref{sec:mag_vs_phase}.
For RealData in the evaluation set, only the \ac{SRMR} results are reported, since a clean reference signal is not available.

\subsection{Results}

\begin{table*}[ht!]
    \centering
    \resizebox{0.8\linewidth}{!}{%
    \begin{tabular}{l r r r r r r r}
        \toprule 
        & & & SimData & & & & RealData \\
        \midrule
        & \textbf{CD} $\downarrow$ & \textbf{LLR} $\downarrow$ & \textbf{PESQ} $\uparrow$ & \textbf{SI-SDR} $\uparrow$ & \textbf{SNR}$_{\textrm{fw}}$(dB)  $\uparrow$ &  \textbf{SRMR} $\uparrow$ & \textbf{SRMR} $\uparrow$ \\
        \midrule 
        \textbf{Reverb} & 3.97 & 0.58 & 1.48 & -10.4 & 3.68 & 3.62 & 3.18 \\
        \textbf{WPE} (1-ch) ~\cite{delcroix2014linear} & 3.74 & 0.52 & 1.72 & - & 4.90 & 4.22 & 3.97 \\
        \textbf{DNN}~\cite{xiao2016speech} & 2.50  & 0.50 & - & - & 7.55 & \textbf{5.77} & 4.36 \\
        \textbf{WRN} ~\cite{ribas2019deep} & 3.59 & 0.47 & - & - & 4.80 & 3.59 & 3.24 \\
        \textbf{TCN+SA} ~\cite{zhao2020monaural} & 2.20 & 0.24 & 2.58 & - & \textbf{13.06} & 5.17 & 5.54 \\
        \textbf{U-Net} ~\cite{ernst2018speech} & 2.50 & 0.40 & - & - & 10.70 & 4.88 & 4.88 \\
        \midrule
        \textbf{S2S+NP} & \textbf{1.95} & \textbf{0.20} & \textbf{2.62} & -9.48 & 12.20 & 4.63 & 6.08 \\
        \textbf{S2S+CP} & 1.72 &  0.18 &  3.09 & 10.86 & 13.75 & 4.92 & - \\
        \textbf{RI2RI (MSE)} & 3.84 & 0.65 & 1.55 & 0.36 & 8.95 & 4.95 & 5.98 \\
        \textbf{RI2RI (SI-SDR)} & 3.57 & 0.60 & 1.67 & 1.40 & 8.86 & 5.20 & 6.72 \\
        \textbf{S2S+RI2RI} & 2.93 & 0.41 & 2.38 & \textbf{1.94} & 10.93 & 4.89 & \textbf{7.49} \\
        \bottomrule 
    \end{tabular}}
    \caption{Average performance of different algorithms on SimData and RealData of REVERB challenge evaluation-set. The upper part presents results of the compared algorithms. The results in the upper part are their reported results. The lower part presents the results of different variants we implemented and the proposed method (in the last row).}
    \label{table:3}
\end{table*}

Table~\ref{table:3} describes the objective measures of the compared algorithms and variants on the simulated dataset (SimData) and on the real recording data (RealData). The upper part of the table is dedicated to the compared methods, while the lower part is dedicated to the different variants of the proposed method.

Focusing on the upper part, it is easy to see that the DNN-based models outperform the classic algorithm. It is also clear that all the methods improves the magnitude related measurements. 
In the lower part of the table, we first note that our S2S+NP model (which is very similar to the model in~\cite{zhao2020monaural}) demonstrates the best magnitude-related results. Interestingly, the same model with the clean phase improved dramatically, the \ac{SI-SDR} results as expected, while still best performing also for the magnitude-based measures. 
 The \ac{RI2RI} (MSE) model  improves the SI-SDR on the one hand, while the \ac{PESQ}, \ac{CD} and the \ac{LLR} measures do not improve so much, on the other hand. Similarly, the \ac{RI2RI} (SI-SDR) model,  even improves the SI-SDR slightly better, while the magnitude-related measures exhibits only marginal improvement. 

Finally, our two-stage proposed method called S2S+RI2RI was tested.  
Evidently, our approach  improves the magnitude and as well as the phase related measurements. While improving the \ac{SI-SDR} in one hand, the \ac{PESQ}, \ac{CD} and the LLR are still competitive with the the best \ac{S2S} models. 
It is worth noting that in the RealData test set, our approach even gained \ac{SOTA} results. 
Sound examples are available in our website.\footnote{\texttt{https://sharongannot.group/audio/}}

\subsection{Ablation study}
\label{Ablation_study}
The proposed architecture comprises two blocks, which are responsible for the magnitude and the phase components of the \ac{STFT}. There are few ways to train such a complex two-stage architecture. First, we found that without pre-training the blocks, namely when only joint training is applied, the algorithm did not converge. Hence, pre-training is required. After pre-training both sub-models separately we test whether a joint fine-tuning  is needed, and if so, do we freeze or not one of the models during the joint training.
Table~\ref{table:ablation} depicts different training options. The first row represents only the pre-training without additional fine-tuning. The last one represents a full joint fine-tuning. Note that in the fine-tuning phase, the active learning models were tuned with their corespondent loss described in Sec.~\ref{sec:pre_train}. 
It is evident that the best results are obtained with the following approach. Consequently ,the final training scheme is obtained with first, pre-training each model separately and in then, freeze the S2S model  and fine-tune the entire architecture jointly.

\begin{table}[ht!]
\vspace{-0.3cm}
 \caption{Average performance of different training combinations on SimData of REVERB challenge evaluation set. The best combination was found with freeze the S2S model for maintain magnitude performance and fine-tuning the \ac{RI2RI} part (after pre-train on clean magnitude) for the best focusing on the phase enhancement task.}
\centering
\resizebox{\linewidth}{!}{%
 \begin{tabular}{l l r r r r}
 \midrule
 \textbf{S2S} & \textbf{RI2RI} & \textbf{LLR} $\downarrow$ & \textbf{CD}  $\downarrow$ & \textbf{PESQ}  $\uparrow$ & \textbf{SI-SDR}  $\uparrow$
\\ \midrule
freeze & freeze  & 0.45 & 3.18 & 2.32 & 0.77 \\
no freeze  & freeze  & 0.48 & 3.33 & 2.35 & 0.28 \\
freeze & no freeze   & \textbf{0.41} &  \textbf{2.93} & \textbf{2.38} & \textbf{1.94} \\
no freeze & no freeze   & 0.67 & 4.27 & 2.26 & 1.64 \\
\midrule 
\end{tabular}}
\label{table:ablation}
\end{table}

\section{Conclusion}

In this paper we presented a new deep learning architecture for enhancing reverberated speech signal. We first showed an analysis study implying that focusing on magnitude or phase solely is not sufficient. Furthermore we demonstrated that some objective measurements are affected by the magnitude and some by the phase.  The proposed method, build to improve both aspects, is built of two sub-models, where one is focused on the magnitude enhancement and the other on the phase enhancement. We described a training scheme to train this architecture. 
Experiments  on the REVERB challenge show consistently improvement in all measurements, and in real dataset our approach is the SOTA. 




\bibliographystyle{IEEEbib}
\bibliography{strings,refs}

\end{document}